# ArEEG_Words: Dataset for Envisioned Speech Recognition using EEG for Arabic Words

Hazem Darwish, Abdalrahman Al Malah, Khloud Al Jallad, Nada Ghneim.

k-aljallad@aiu.edu.sy, n-ghneim@aiu.edu.sy
201910477@aiu.edu.sy, 201910306_@aiu.edu.sy
Department of Information and Communication Engineering,
Arab International University, Daraa, Syria.

## Abstract

Brain-Computer-Interface (BCI) aims to support communication-impaired patients by translating neural signals into speech. A notable research topic in BCI involves Electroencephalography (EEG) signals that measure the electrical activity in the brain. While significant advancements have been made in BCI EEG research, a major limitation still exists: the scarcity of publicly available EEG datasets for non-English languages, such as Arabic. To address this gap, we introduce in this paper ArEEG_Words dataset, a novel EEG dataset recorded from 22 participants with mean age of 22 years (5 female, 17 male) using a 14-channel Emotiv Epoc X device. The participants were asked to be free from any effects on their nervous system, such as coffee, alcohol, cigarettes, and so 8 hours before recording. They were asked to stay calm in a clam room during imagining one of the 16 Arabic Words for 10 seconds. The words include 16 commonly used words such as up, down, left, and right. A total of 352 EEG recordings were collected, then each recording was divided into multiple 250ms signals, resulting in a total of 15,360 EEG signals. To the best of our knowledge, ArEEG_Words data is the first of its kind in Arabic EEG domain. Moreover, it is publicly available for researchers as we hope that will fill the gap in Arabic EEG research.

**Keywords:** EEG, Arabic words EEG Dataset, Brain-computer-Interface BCI

---

## 1. Introduction:

An important area of computer science that examines the relationship between people and computers is called human-computer interaction, or HCI.

Brain-Computer-Interface (BCI) is one of the HCI fields that has seen a lot of researches in the last few years. Researchers are using artificial intelligence (AI) to interpret brainwaves, measured through electroencephalography (EEG), to understand what people are thinking. Recent advancements in BCI have shown promise in recognizing mental activities like imagining speech. However, using EEG signals to directly recognize spoken language through BCI is still a relatively new area of research.

Arabic language is an important and rich language spoken by over 400 million people worldwide, and it is the official language in many countries. However, Arabic Language is still one of the most challenging low-resource languages.

In this paper, we present our method of creating ArEEG_Words, an EEG dataset that contains signals of some Arabic words. We have reviewed the models used in the literature to classify the EEG signals, and the available datasets for English.

The main contribution of this paper is creating a dataset for EEG signals of all Arabic words that will be publicly available for researchers[1]. To the best of our knowledge, ArEEG_Words is the first Arabic EEG dataset for words.

The rest of the paper is organized as follows: Related works are presented in Section 2, Section 3 specifies the data collection method and the resulting dataset, and finally, section 4 gives the conclusion.

2. **Related Works:**

In 1924, Hans Berger [1] published the first paper about EEG classification, which was used later to develop EEG-based brain-computer interface (BCI). The main idea that Berger proposed a method that make use of EEG signals' frequencies to classify EEG signals into different categories.

There are several EEG publicly available datasets collected using different EEG devices, for various languages and covered a wide range of recognized units, including digits, characters, directions, audio, images, and words. This paper will specifically focus on word-based EEG datasets.

Each study involved a varying number of participants, ranging from 3 to 29, with most being healthy adults. Data collection methods varied significantly. Participants might be shown objects for a certain duration, then asked to imagine the object, hear the imagined word mentally, or a combination of these. In some studies, participants were required to be free from any nervous system stimulants like caffeine, alcohol, or nicotine.

In [2] Pradeep Kumar et al. developed a system for recognizing imagined speech based on EEG signals. They collected data from 23 participants using an EPOC+ device where EEG signals has been recorded using 30 text and non-text class objects: letters, images and digits being imagined by multiple participants. To collect this data, participants were shown an object on a screen, then closed their eyes and imagined it for 10 seconds. After a 20-second rest period, they were shown the next object. A total of 230 EEG samples were collected for each object category and participant, resulting in a dataset of 230*30 *23 samples.

Nicolás Nieto in [3] created a publicly available EEG-based BCI dataset for inner speech recognition using EEG 16, 17, 18, 19, 20, and 21 sensors on 10 participants (6 male, 4 female). Each participant first heard one of 4 directions (up, down, left, and right) through a loudspeaker (average length: 800ms ± 20). Then, a first cross was shown on a screen (1500ms after trial onset) for 1000ms, indicating that the participant had to imagine hearing the word. Finally, a second cross was shown on the screen (3000ms after trial onset) for 1500ms, indicating that the participant had to repeat out loud the word. The words were chosen to maximize the variability of acoustic representations, semantic categories, and the number of syllables while minimizing the variability of acoustic duration.

---

[1] Darwish, Hazem; Almalah, Abdalrahman; Al Jallad, Khloud; Ghneim, Nada; (2024), "ArEEG_Words", Mendeley Data, V1, doi: 10.17632/7m472ykkx7.1, url: https://data.mendeley.com/datasets/7m472ykkx7.1

In [4], EEG and deep learning were used to classify imagined speech commands into four direction classes (up, down, left, right). Data were collected from four participants using Unicorn Hybrid Black+ headset eight EEG sensors. Data contained different frequency bands with different amplitude ranges. Each trial began by speaking command. After 10-12 seconds, participant silently imagined and spoke the command for 60 seconds, followed by a 10-second recording pause. Each participant provided 100 recordings for each of the four commands, resulting in a total of 400 recordings.

In [5], seven people participated in the experiment which was done in a poorly light room where they minimized body and eye movements. EEG data was recorded using a 128-channel Geodesic Sensor Net with an ANT-128 amplifier at a 1024 Hz sampling rate. Participant's task was to imagine speaking one of two syllables, (/ba/ or /ku/), in one of three rhythms. The first rhythm has the time structure {| 1.5 | 1.5 |}, the second has the structure {| 0.75 | 0.375 | 0.375 | 0.75 | 0.375 | 0.375 |}, and the third has the structure {| 0.5 | 0.5 | 0.5 | 0.5 | 0.5 | 0.5 |}; the vertical lines "|" represent the expected times of imagined syllable production onset, while the numbers indicate the time intervals in seconds between imagined syllables. Each experiment began with a 4.5-second cue period, where a syllable cue (/ba/ or /ku/) was shown for 0.5 seconds, followed by rhythmic cues. After a 1-second silence for baseline estimation, participants imagined speaking the syllable for 6 seconds. Each person completed 720 experiments across three rhythms and two syllables.

As for German language, in their study "Imagined speech event detection from electrocorticography and its transfer between speech modes and subjects," Aurélie de Borman et al. in [6] analyzed ECoG signals from 16 participants with subdural implants to detect imagined speech. Three speech tasks were given to participants: perform, perceive, and imagine speech. The Electrocorticography ECoG signals were recorded in seven frequency bands, from delta (0.5–4 Hz) to high-gamma (70–120 Hz). In each experiment, participants were given a short sentence to memorize it, and then perform one of the speech tasks. The short sentences imagined by the participants in the study were selected from the LIST database [7], which contains Dutch sentences. Sentences were shown randomly in one of two tasks vocalizing and listening.

Table 1 shows comparisons between EEG datasets.

| Type | Study | # Samples | # Participants | Age | Device | Frequencies | # Letter, Words & Images | Recording time |
|---|---|---|---|---|---|---|---|---|
| Digits & chars | [2] | 230 EEG sample for each per category for participant | 23 | 15-40 | 14 Channels EMOTIV EPOC | 2048Hz down-sampled 128Hz | 10 objects of each class (digits, alphabets, images) | 10 seconds |
| Directions | [3] | from 18 to 24 trials for each | 10 | 31 in average | 16, 17, 18 19. 20, and 21 sensors EEG headcap and the external electrodes | 1024Hz | 4 words | 5.8 seconds |
| | [4] | / | 4 | 32 | Unicorn Hybrid Black+ | 250Hz | 4 words | 60 seconds |
| Vowels | [8] | 100 sample | 3 | 26-29 | 128 Channels Electrical Geodesics | 1024Hz | 2 vowels | 2 seconds |
| | [9] | / | 5 | 21-24 | 19 Channels | / | 5 vowels | / |
| Directions & Select | [10] | 33 epoch each subject | 27 | / | 14 Channels EMOTIV EPOC | 128Hz | 5 words | / |
| | [11] | 33 epoch each subject | 27 | / | 14 Channels EMOTIV EPOC | 128Hz | 5 words | / |
| Syllables | [5] | 120 trials | 7 | / | 128-channel Geodesic Sensor Net | 1024 Hz. | two syllables /ba/ and /ku/ | 10.5 seconds |
| Words | [6] | 60 trials per speech mode | 16 | / | Subdural electrodes, SD LTM 64 Express amplifier | 256 Hz | 20 Dutch words | 10.5 seconds |

*Table 1 Comparison Between EEG Datasets*

## 3. ArEEG_Words Dataset

### 3.1. Dataset Collection

EGG data was recorded using the Emotiv EPOC X headset from 22 adult participants (17 males, 5 females) who were educated at university level. Participants were told to maintain calm with clear thoughts throughout the data collection process. Additionally, they were advised not to consume caffeine, alcohol, or smoke before 8 hours of the recordings to prevent any impact on their nervous systems. Each participant spent approximately about one hour in a comfortable room wearing the headset and focusing on recording their thoughts related to a set of 16 Arabic words presented on individual slides. Figure 1 shows all the used slides. The order of the word slides was randomized to prevent anticipation. Following the 10-second display of each word, participants were given 10 seconds to envision the word with their eyes closed. A 20-second break between recordings allowed participants to reset their mental state. Periodically during the session, participants were asked if they maintained focus while imagining the word, ensuring compliance with the study protocol. A total of 352 EEG recordings were collected using this methodology, with each recording divided into multiple 250ms signals for analysis, resulting in a total of 15,360 EEG signals.

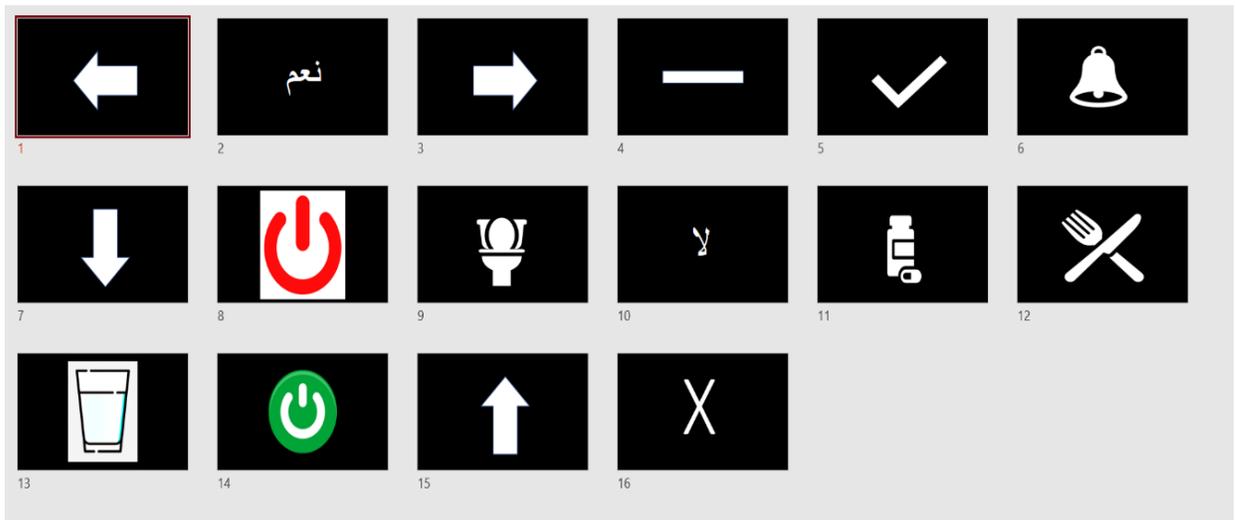

*Figure 1 Words used in this study*

### 2.1. Headset:

In our framework, a wireless neuro-headset known as Emotiv EPOC X has been used for the acquisition of the envisioned brain signals. For recording brain signals, this device incorporates 14 channels namely AF3, F7, F3, FC5, T7, P7, O1, O2, P8, T8, FC6, F4, F8, AF4, which are placed over the scalp of the user according to the International 10-20 system as shown in Figure 2, where two reference electrodes, i.e., CMS and DRL, are positioned above the ears.

We initially captured the brain waves at a frequency of 2048 Hz and later down-sampled them to 128 Hz per channel. The captured brain waves were sent to the computing device through Bluetooth technology.

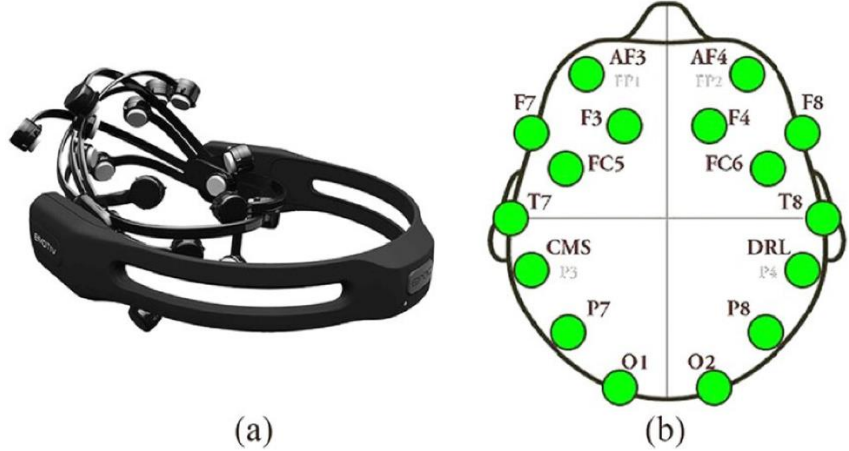
*Figure 2 EEG headset and International 10-20 system*

## 2.2. Statistics about our dataset

Twenty-two Arabic-native-speakers contributed in our experiments. Table 2 provides information about all participants. The average age is 22.5.

| Participant No. | Age | Gender |
|---|---|---|
| Par.1 | 24 | Female |
| Par.2 | 25 | Male |
| Par.3 | 23 | Female |
| Par.4 | 25 | Male |
| Par.5 | 23 | Male |
| Par.6 | 23 | Male |
| Par.7 | 17 | Male |
| Par.8 | 24 | Male |
| Par.9 | 25 | Male |
| Par.10 | 22 | Male |
| Par.11 | 22 | Male |
| Par.12 | 22 | Male |
| Par.13 | 22 | Male |
| Par.14 | 24 | Male |
| Par.15 | 23 | Female |
| Par.16 | 22 | Male |
| Par.17 | 22 | Female |
| Par.18 | 21 | Male |
| Par.19 | 23 | Male |
| Par.20 | 23 | Female |
| par.21 | 19 | Male |
| Par.22 | 22 | Male |

*Table 2 Participants' information*

The majority of participants are male (70%) with ages between 16 and 64 as shown in Figure 3. Moreover, the majority are high school students, As shown in Figure 4.

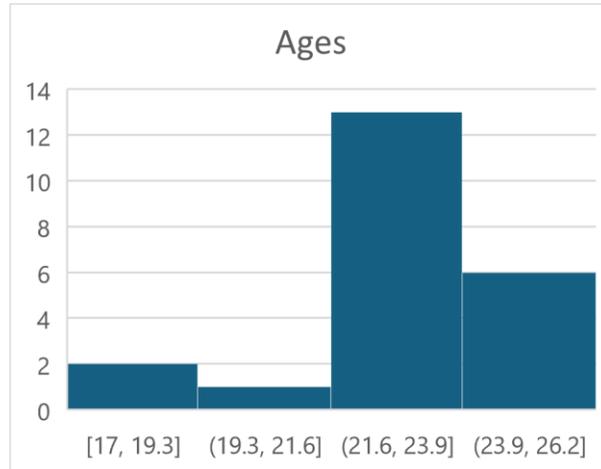

Figure 3 Participants Ages Distributions

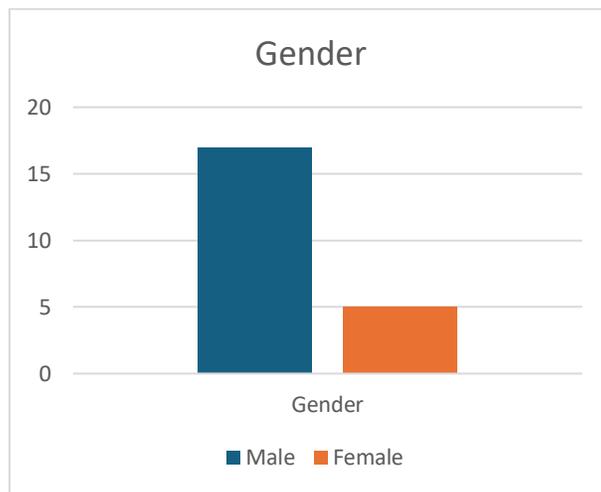

Figure 4 Participants Gender Ratio

4. **Conclusion**

This research aims to fill a gap in Arabic Brain-Computer Interface (BCI) research by creating a publicly available dataset of EEG signals corresponding to specific Arabic words, called ArEEG_Words. This dataset includes data for 16 Arabic words collected from 22 participants, primarily young adults with mean age of 22 years (5 female, 17 male). As for future work, we are willing to create baseline models using deep learning to predict Arabic words from the EEG signals. Furthermore, as the importance of a richer data pool, we think of expanding ArEEG_Words in future by adding more words and more participants. By making this dataset public, we aim to accelerate advancements in BCI technology, particularly for Arabic native individuals with communication disabilities.

# Reference

[1]   Lorainne Tudor Car; Mario Tudor; Katarina Ivana Tudor, "Hans berger The history of electroencephalography," *Acta medica Croatica: časopis Hravatske akademije medicinskih znanosti,* p. 8, 2005.

[2]   Pradeep Kumar, Rajkumar Saini, Partha Pratim Roy, Pawan Kumar Sahu, and Debi Prosad Dogra., "Envisioned speech recognition using EEG sensors.," *Personal and Ubiquitous Computing.,* vol. 15, 2017.

[3]   Nicolás Nieto; Victoria Peterson. Hugo Leonardo Rufner; Juan Esteban Kamienkowski; Ruben Spies, "Thinking out loud, an open-access EEG-based BCI dataset for inner speech recognition," *Scientific data,* p. 17, 2022.

[4]   Mokhles M. Abdulghani; Wilbur L. Walters; Khalid H. Abed, "Imagined Speech Classification Using EEG and Deep Learning," *bioengineering,* p. 15, 2023.

[5]   Siyi Deng; Ramesh Srinivasan., " EEG classification of imagined syllable rhythm using Hilbert spectrum methods.," *Journal of Neural Engineering,* p. 25, 2010.

[6]   Aurélie de Borman; Benjamin Wittevrongel; Ine Dauwe; , Evelien Carrette; Alfred Meurs; Dirk Van Roost; Paul Boon; & Marc M. Van Hulle , "Imagined speech event detection from electrocorticography and its transfer between speech modes and subjects," *communications biology,* p. 12, 2024.

[7]   A. van Wieringen and J. Wouters, "LIST and LINT: sentences and numbers for quantifying speech understanding in severely impaired listeners for Flanders and the Netherlands," *Int J. Audio,* vol. 47, p. 348–355, 2008 .

[8]   Charles S. DaSalla, Hiroyuki Kambara, Makoto Sato, Yasuharu Koike., "Single-trial classification of vowel speech imagery using common spatial patterns.," *Neural Networks.,* p. 70, 2009.

[9]   Mariko Matsumoto; Junichi Hori, "Classification of silent speech using support vector machine and relevance vector machine," *Applied Soft Computing,* p. 20, 2014.

[10] Alejandro A Torres-Garc´ıa; Carlos A. Reyes-Garc´ıa; Luis Villasenor-Pineda; Gregorio Garc´ıa-Aguilar, "Implementing a Fuzzy Inference System in a Multi-Objective EEG Channel Selection Model for Imagined Speech Classification," *Expert Systems With Applications,* p. 41, 2016.

[11] Luis Alfredo Moctezuma; Alejandro A. Torres-Garc´ıa; Luis Villasenor-Pineda; Maya Carrillo-Ruiz., "Subjects Identification using EEG-recorded Imagined Speech," *Expert Systems With Applications,* p. 26, 2018.

[12] A. A. Torres-García, C. A. Reyes-Garcia and L. Villaseñor-Pineda, "Toward a silent speech interface based on unspoken speech," in *Proceedings of BIOSIGNALS 2012 (BIOSTEC)*, Algarve, Portugal, 2018.

[13] A. Gramfort, M. Luessi, E. Larson, D. A. Engemann, D. Strohmeier, C. Brodbeck, L. Parkkonen, M. S. Hämäläinen, , "MNE software for processing MEG and EEG data," *NeuroImage,* p. 86, 2014.